\documentclass[a4paper]{jpconf}
\usepackage{graphicx}
\begin{document}
\begin{flushright}
IFIC 09-49
\end{flushright}

\title{Probing the Majorana nature of the neutrino with 
neutrinoless double beta decay
}

\author{Stefano Morisi}

\address{  Institut de F\'{\i}sica Corpuscular --
  C.S.I.C./Universitat de Val{\`e}ncia \\
  Edificio Institutos de Paterna, Apt 22085, E--46071 Valencia, Spain
}

\ead{morisi@ific.uv.es}

\begin{abstract}
Neutrinoless double beta decay ($0\nu  \beta \beta$) is the only experiment that could probe the Majorana
nature of the neutrino. Here we study the theoretical implications of $0\nu  \beta \beta$ for 
models yielding tri-bimaximal lepton mixing like $A_4$ and $S_4$. 

\end{abstract}

\section{Introduction}
Neutrinoless double beta decay ($0\nu  \beta \beta$) is 
the only experiment that can probe the Majorana Nature of the neutrino.
$0\nu  \beta \beta$ is a $\Delta L =2 $ lepton number  violating decay
that can occur if the neutrino is a Majorana particle. 
For an introduction see for instance \cite{Hirsch:2006tt}.
The generic effective
Majorana mass term can be write as 
\begin{equation}\label{l5d}
\mathcal{L}_{d5}= \frac{\lambda_{ij}^\nu}{\Lambda} L_i\phi L_j\phi,
\end{equation}
where $\phi$ is the standard model higgs doublet and $\Lambda$ is 
an effective scale. When $\phi$ takes a vev $\langle \phi \rangle =v$, the operator in eq.\,(\ref{l5d})
gives the Majorana mass term $m_{ij}^\nu=\lambda_{ij}^\nu v^2/\Lambda$.
In general $m^\nu$ is an arbitrary 3 by 3 symmetric complex
matrix. 
It has been observed \cite{Harrison:2002er} that neutrino data are in
very good agreement with zero a rector angle $\sin \theta_{13}=0$, 
maximal atmospheric angle $\sin^2\theta_{23}=1/2$ and large but not maximal 
solar angle  $\sin^2\theta_{12}=1/3$. 
A Leptonic mixing mixing matrix with
such a values is called tri-bimaximal (TBM). The most generic neutrino mass 
matrix yielding to TBM mixing is invariant under exchanged the second family 
with the third family ($\mu \leftrightarrow \tau$  symmetry) and the sum
of the elements each row of $m_\nu$ is constant, namely 
$\sum_j m^\nu_{1j}=\sum_j m^\nu_{2j}=\sum_j m^\nu_{3j}$. We call a neutrino 
mass matrix of such type tri-bimaximal form mass matrix.

We can extend the standard model by means of a flavor symmetry $G_f$ that
yields a  tri-bimaximal form mass matrix. Examples are the group
of even permutation of four objects $A_4$ \cite{a4,Altarelli:2005yp} 
or the group of all the 
permutation of four objects $S_4$ \cite{Lam:2008rs,Bazzocchi:2008ej}. We observe that $A_4$ is the 
smallest non-Abelian group with a triplet representation allowing to arrange in one 
multiplet the three families.
For now on we focus on $A_4$ flavor symmetry. We assume three Higgs doublets 
transforming as a triplet $(\phi_1,\phi_2, \phi_3 )\sim 3$
and the lepton doublets $L_e,L_\mu,L_\tau$, transforming as a triplet of $A_4$.
The most generic $A_4$ invariant dimension five operators are given as
\begin{eqnarray}
\mathcal{L}_{5d}&=&
\frac{c_1}{\Lambda} (L_i\phi)_1 (L_j\phi)_1+
\frac{c_2}{\Lambda} (L_i\phi)_{1'} (L_j\phi)_{1''}+
\frac{c_3}{\Lambda} (L_i\phi)_3 (L_j\phi)_3+\nonumber\\
&+&\frac{c_4}{\Lambda} (L_iL_j)_1 (\phi \phi)_1+
\frac{c_5}{\Lambda} (L_iL_j)_{1'} (\phi \phi)_{1''}+
\frac{c_5'}{\Lambda} (L_iL_j)_{1''} (\phi \phi)_{1'}+
\frac{c_6}{\Lambda} (L_iL_j)_3 (\phi \phi)_3.\label{S4}
\end{eqnarray}
The resulting neutrino mass matrix is of tri-bimaximal form if and only
if $c_5=c_5'=0$. 
Such a condition can naturally be explained
by adding Higgs doublets or flavons carrying
extra Abelian symmetries, see for instance \cite{Altarelli:2005yp}.

Differently
the most generic $S_4$ invariant dimension five operators 
\begin{eqnarray}
\mathcal{L}_{5d}&=&
\frac{\lambda_1}{\Lambda} (L_i\phi)_1 (L_j\phi)_1+
\frac{\lambda_2}{\Lambda} (L_i\phi)_{2} (L_j\phi)_{2}+
\frac{\lambda_3}{\Lambda} (L_i\phi)_{3_1} (L_j\phi)_{3_1}+
\frac{\lambda_3'}{\Lambda} (L_i\phi)_{3_2} (L_j\phi)_{3_2}+\nonumber\\
&+&\frac{\lambda_4}{\Lambda} (L_iL_j)_1 (\phi \phi)_1+
\frac{\lambda_5}{\Lambda} (L_iL_j)_{2} (\phi \phi)_{2}+
\frac{\lambda_6}{\Lambda} (L_iL_j)_{3_1} (\phi \phi)_{3_1}
,\label{S4}
\end{eqnarray}
give neutrino mass matrix of tri-bimaximal form. In $S_4$ we do 
not need to introduce
extra scalars or add symmetries in order to have tri-bimaximal in the 
neutrino sector. The resulting neutrino mass matrix is diagonalized 
by the tri-bimaximal matrix.

\begin{center}
\begin{figure}[t]
\begin{minipage}{16pc}
\hskip5.mm
\includegraphics[width=14pc]{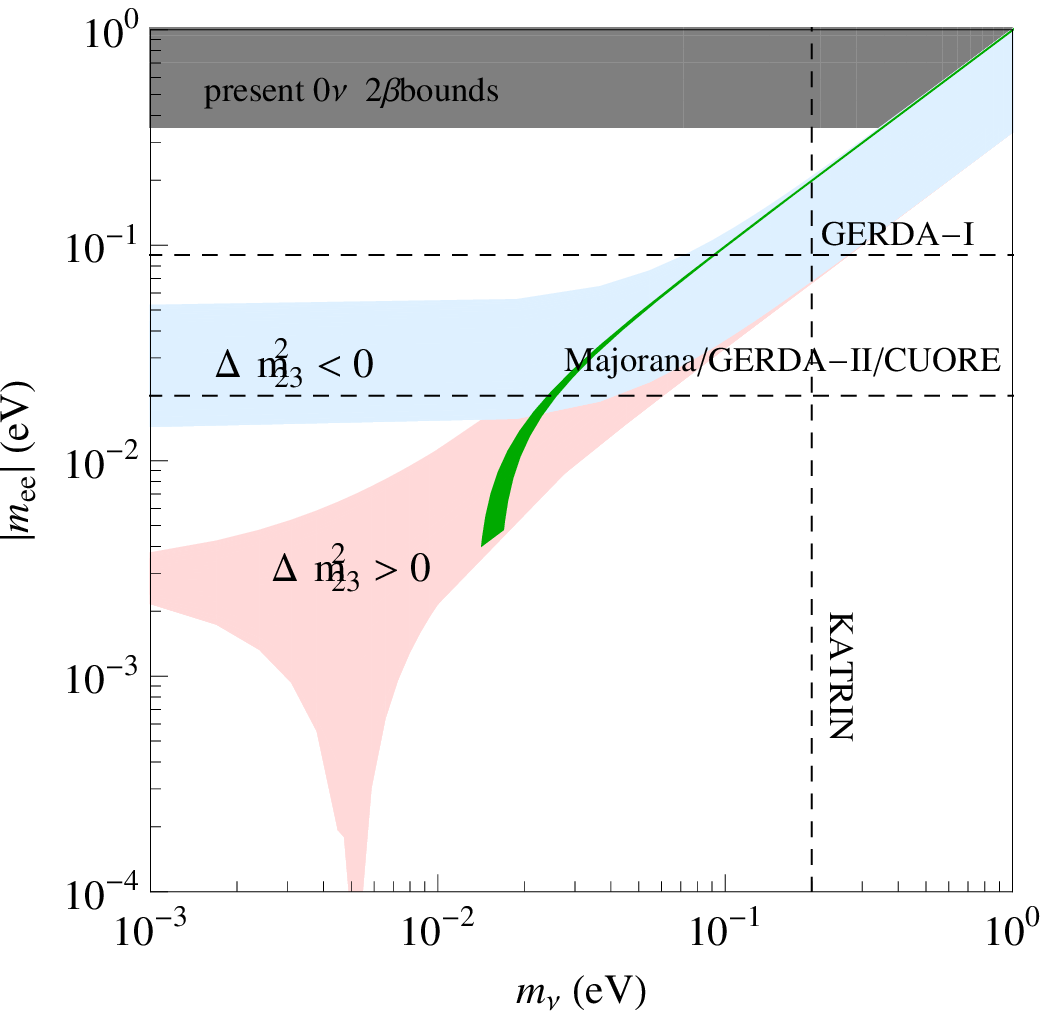}
\caption{\label{label1}Allowed range of $m_{ee}$ as a function of
the lightest neutrino mass for $A_4$ based model.}
\end{minipage}\hspace{2pc}%
\begin{minipage}{16pc}
\hskip5.mm
\includegraphics[width=14pc]{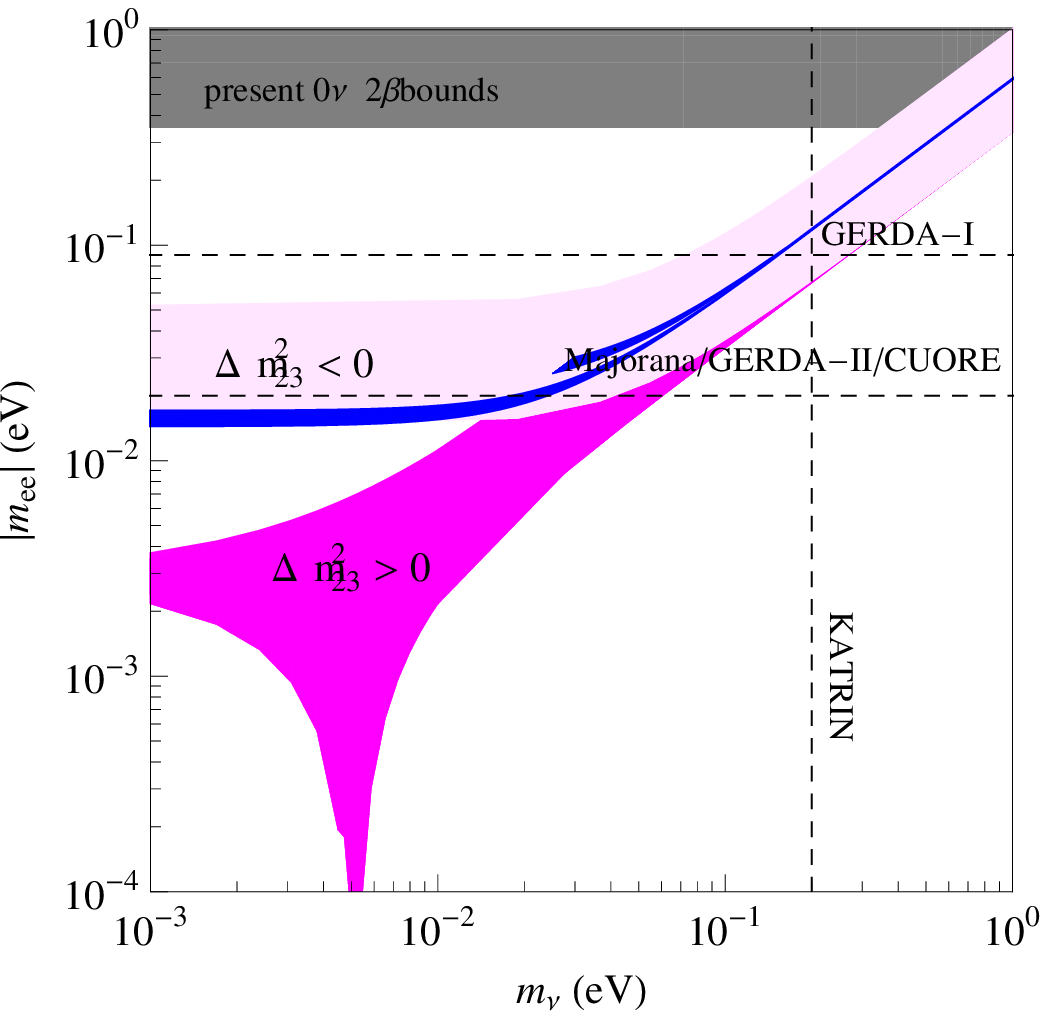}
\caption{\label{label2}Allowed range of $m_{ee}$ as a function of
the lightest neutrino mass for $S_4$ based model.}
\end{minipage} 
\end{figure}
\end{center}

Now we study the theoretical implications for $0\nu  \beta \beta$ 
in models based on the two flavor groups $A_4$ and $S_4$. 
The allowed ranges of
$m_{ee}$ as a function of the lightest neutrino are presented in 
Figures\,(\ref{label1})\,and\,(\ref{label2}). The two bands  correspond
to the normal hierarchy and the inverse hierarchy
with $\sin^2\theta_{12}=1/3$, $\sin^2\theta_{23}=1/2$, $\sin^2\theta_{13}=0$ 
(tri-bimaximal). We call this bands the {\it tri-bimaximal region} for $0\nu\beta\beta$.
In order to simplify the discussion and to show the difference between
$A_4$ and $S_4$ based models, we assume $c_1=c_2=c_3=0$ 
and $\lambda_1=\lambda_2=\lambda_3=\lambda_3'=0$. This could be the case
if the dimension five operators arise from type-II seesaw mechanism only.
We have already said that 
$A_4$ gives tri-bimaximal mixing if $c_5=c_5'=0$. 

If $\lambda_5=0$
$A_4$ and $S_4$ are phenomenologically equivalent since 
the operators proportional to $c_4$ and $c_6$ correspond
 to the operators proportional to 
$\lambda_4$ and $\lambda_6$ respectively.

If $\lambda_5\ne 0$, then $A_4$ and $S_4$ are not phenomenologically equivalent. 
For instance the $S_4$ allowed range  for $m_{ee}$ is the full tri-bimaximal region. 
When  $\lambda_4=0$ 
the allowed $m_{ee}$ ranges for $A_4$ and $S_4$ are given respectively in 
Figures\,(\ref{label1})\,and\,(\ref{label2}). 

In particular, in Figure\,(\ref{label1}) we see that $A_4$ is compatible only with normal hierarchy. 
The $A_4$ case has a lower bound with $m_{ee}>3 \cdot 10^{-3}$, see \cite{Altarelli:2007gb}. 
In Figure\,(\ref{label2}) we see that $S_4$ is compatible only with inverse hierarchy.

\cite{Hirsch:2008rp} and \cite{Hirsch:2009mx} have studied $0\nu\beta\beta$ 
implications  assuming respectively type-I and inverse/linear seesaw mechanisms with $A_4$ flavor symmetry.
An effective neutrino mass model with $S_4$ flavor symmetry has been studied in \cite{Bazzocchi:2009pv}. 
In \cite{Bazzocchi:2009da}  a model with type-I seesaw with $S_4$ flavor symmetry has been studied.

Note that for values of the lightest mass bigger than $10^{-1}$ eV, namely when 
neutrino are quasi degenerate, $A_4$ and $S_4$ allowed ranges are well 
separated. Therefore it could be in principle possible to distinguish
$A_4$ from $S_4$ in the quasi degenerate region.

\vskip2.mm

\begin{center}
\begin{figure}[h!]
\hskip50.mm
\includegraphics[width=14pc]{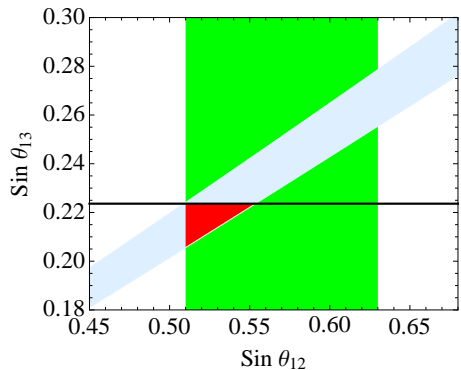}
\caption{\label{label3}Relation between solar and reactor angles for
normal hierarchy when $m_{\nu_1}=m_{ee}=0$.}
\end{figure}
\end{center}

In case of inverse hierarchy there is a lower limit for $m_{ee}$. In 
\cite{Hirsch:2006tt}  the dependence of such a limit
as a function of the solar angle has been studied. However, 
in case of normal hierarchy
in general there is no lower limit. 
If $m_{\nu_1}=m_{ee}=0$ we have that the lower limit depends from the values
of the solar and reactor angles, 
\begin{equation}\label{b1}
\sin^2\theta_{12}\cos^2\theta_{13}e^{i\alpha}\sqrt{\Delta m_{12}^2}+
\sin^2\theta_{13}e^{i\beta}\sqrt{\Delta m_{13}^2}=0.
\end{equation}
This equation gives a relation between the solar and the reactor angles
and the allowed region is represented in Figure\,(\ref{label3}) 
with the blue band. The small red region is the range allowed by 
the data at $3\,\sigma$ and compatible with $m_{\nu_1}=m_{ee}=0$, namely eq.\,(\ref{b1}).

\section*{Acknowledgments}

We thank Martin Hirsch for useful
comments. This work was supported by the Spanish grant
FPA2008-00319/FPA and PROMETEO/2009/091.

\smallskip

\end{document}